# BRIDGING THE MOBILE TRUST GAP: A ZERO TRUST FRAMEWORK FOR CONSUMER-FACING APPLICATIONS

BY ALEXANDER TABALIPA

k9.magnok@gmail.com

https://orcid.org/0009-0000-0023-4980

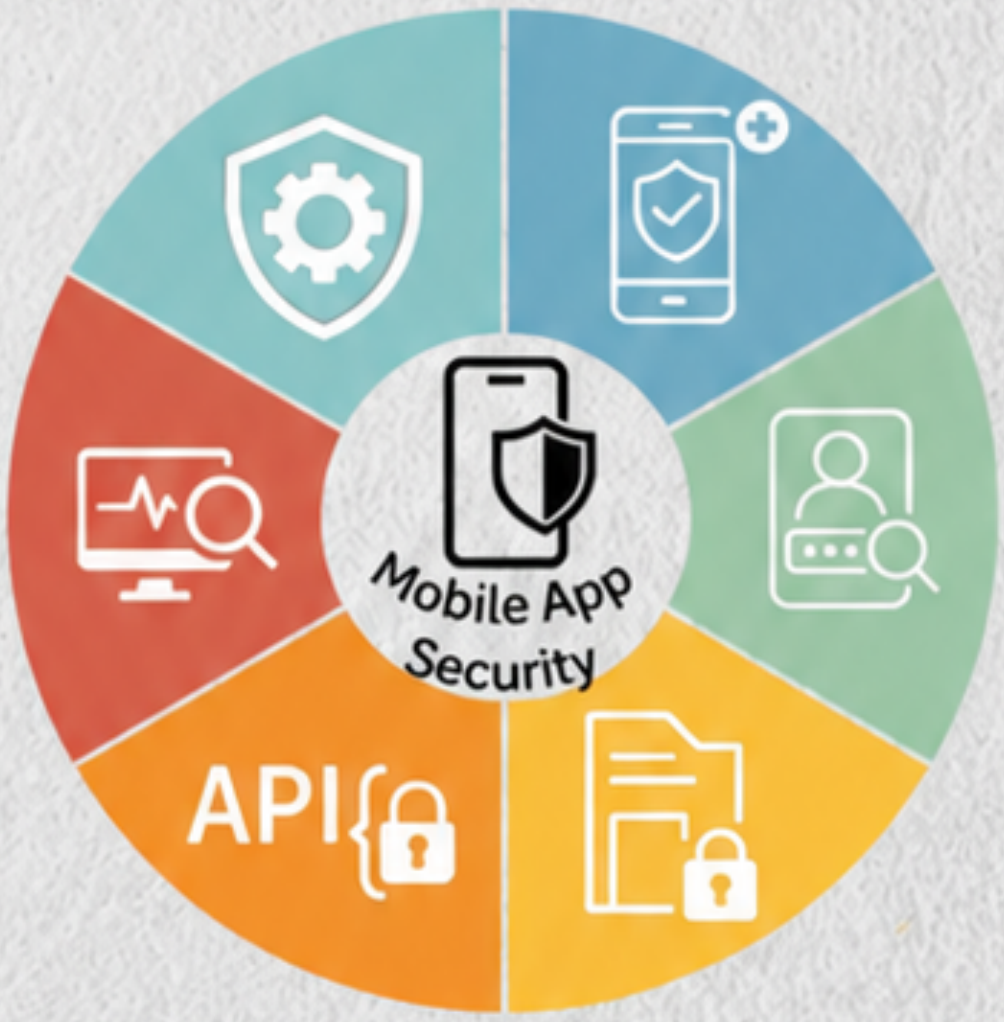

WORKING PAPER – VERSION 2.0

# Bridging the Mobile Trust Gap: A Zero Trust Framework for Consumer-Facing Applications

**Alexander Tabalipa**

12 June 2026

K9.magnok@gmail.com

https://orcid.org/0009-0000-0023-4980

***Working Paper – Version 2.0***

# Abstract

1

2

Zero Trust Architecture (ZTA) has become a widely adopted model for securing enterprise environments, promoting continuous verification and minimal trust across systems. However, although established Zero Trust guidance recognises unmanaged devices and mobile workloads, it provides limited prescriptive detail for assessing application integrity and using runtime risk signals to govern sensitive operations in consumer-facing mobile applications. Existing frameworks from NIST and CISA establish principles for continuous verification, device posture assessment, and contextual access control, but primarily address enterprise resource access and organisational security maturity. This paper addresses the resulting operational gap by proposing an integrated Zero Trust model for mobile applications operating on user-controlled devices. Using a design science methodology, the study introduced a six-pillar framework that supports runtime enforcement of trust through controls including device integrity, user identity validation, data protection, secure application programming interface (API) usage, behavioural monitoring, and live application protection. Each pillar was mapped to relevant regulatory and security standards to support compliance. A phased implementation roadmap and maturity assessment model were also developed to guide adoption across varying organisational contexts. The proposed model offers a practical and standards-aligned approach to securing mobile applications beyond pre-deployment controls, aligning real-time enforcement with Zero Trust principles. This contribution expands the operational boundaries of ZTA and provides organisations with a deployable path to reduce fraud, enhance compliance, and address emerging mobile security challenges. Future research may include empirical validation of the framework and cross-sector application testing.

# Content

## 1.0 Introduction

In today’s digital economy, mobile applications are no longer auxiliary channels but have become the primary interface for customer engagement, transaction execution, and service delivery. More than 90% of mobile time is spent in apps rather than on mobile web browsing (GWI, 2024), with users spending an average of 4 hours and 37 minutes per day on mobile devices (DataReportal, 2025). For many organisations, these applications have become the primary and often the only touchpoint with users. While this shift provides greater convenience and reach, it also introduces significant security risks. Mobile applications operate in untrusted, decentralised environments, often outside the scope of traditional enterprise security tools and policies. As such, they have become preferred vectors for fraud, impersonation, and data compromise.

At the same time, Zero Trust Architecture (ZTA) has emerged as a widely adopted paradigm to secure modern information systems. Defined by the principle of ‘never trust, always verify’, Zero Trust frameworks have reshaped security investments across sectors, with global spending expected to exceed $78 billion by 2029 (MarketsandMarkets, 2024). Foundational guidance from NIST (2020a) and CISA (2023) establishes principles for continuous verification, device posture assessment, and dynamic access control, including scenarios involving personally owned devices and applications running on mobile platforms. However, these publications primarily address enterprise resource protection and organisational Zero Trust maturity, rather than prescribing how consumer-facing mobile applications should establish runtime assurance and translate application integrity signals into enforceable trust decisions on devices outside the organisation's administrative control.

This paper argues that a critical blind spot exists in the application of Zero Trust principles to mobile runtime environments. Recent fraud intelligence highlights the evolving nature of mobile security risks. According to LexisNexis Risk Solutions (2026), mobile applications accounted for 83% of mobile transactions observed within its Digital Identity Network in 2025. Although the overall mobile app attack rate declined by 56% year-over-year to 0.4%, the attack rate targeting password-reset interactions through mobile applications increased by 66%, demonstrating that specific high-risk interactions continue to face significant threats. Although the global Zero Trust security market was estimated at $36.5 billion in 2024 (MarketsandMarkets, 2024), the extent to which these investments specifically address runtime protection in consumer-facing mobile applications remains difficult to quantify. Existing market analyses classify Zero Trust investments across broad categories, including network, endpoint, application, and identity security, without separately identifying consumer mobile runtime protection. This limitation in market segmentation highlights the need to examine mobile runtime assurance as a distinct architectural concern, rather than assuming that existing Zero Trust investments adequately address this attack surface. Threat actors have adapted swiftly, exploiting runtime vulnerabilities, session weaknesses, and API exposures to circumvent conventional controls. As a result, organisations face mounting financial, regulatory, and reputational risks.

Figure 1 illustrates the proposed operational trust gap between established enterprise Zero Trust controls and the execution environment of consumer-facing mobile applications. Although existing Zero Trust architectures incorporate device posture assessment and contextual access control, they do not prescribe a unified mechanism for continuously evaluating application runtime integrity and incorporating that evidence into sensitive API and transaction authorisation decisions on user-controlled devices.

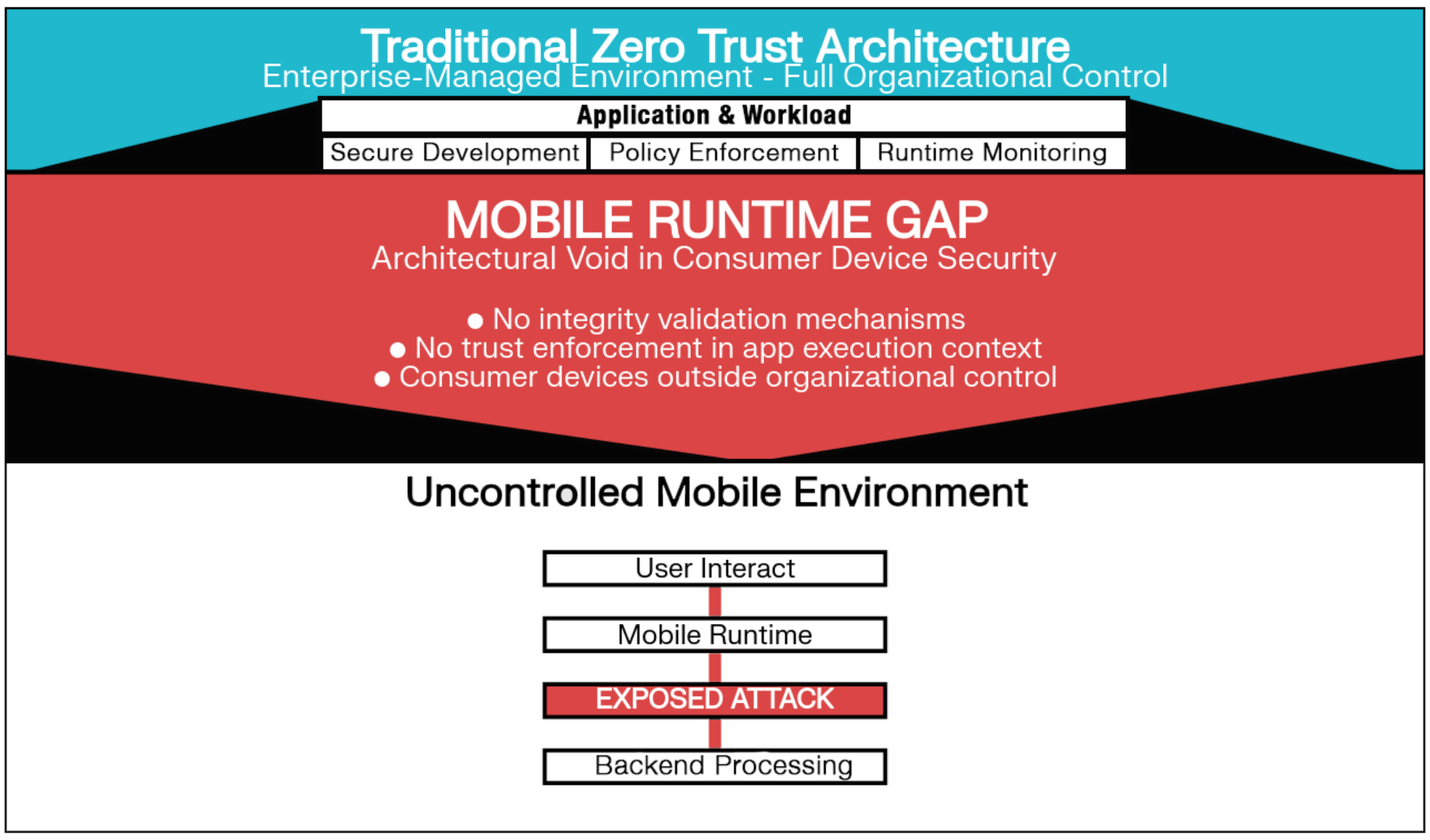

**Figure 1: The Mobile Trust Gap**

To address this gap, we introduce a mobile-specific Zero Trust framework designed to operationalise trust at the runtime layer – where user interaction, threat activity, and application behaviour converge. The proposed six-pillar model integrates architectural guidance from the National Institute of Standards and Technology (NIST) SP 800-207 (NIST, 2020a) and maturity principles from CISA's Zero Trust Maturity Model (CISA, 2023), while incorporating mobile security standards from OWASP MASVS (2024). By aligning Zero Trust enforcement with live mobile conditions, the framework enables continuous identity verification, device integrity assessment, secure data handling, authenticated API usage, behavioural monitoring, and in-app runtime protection.

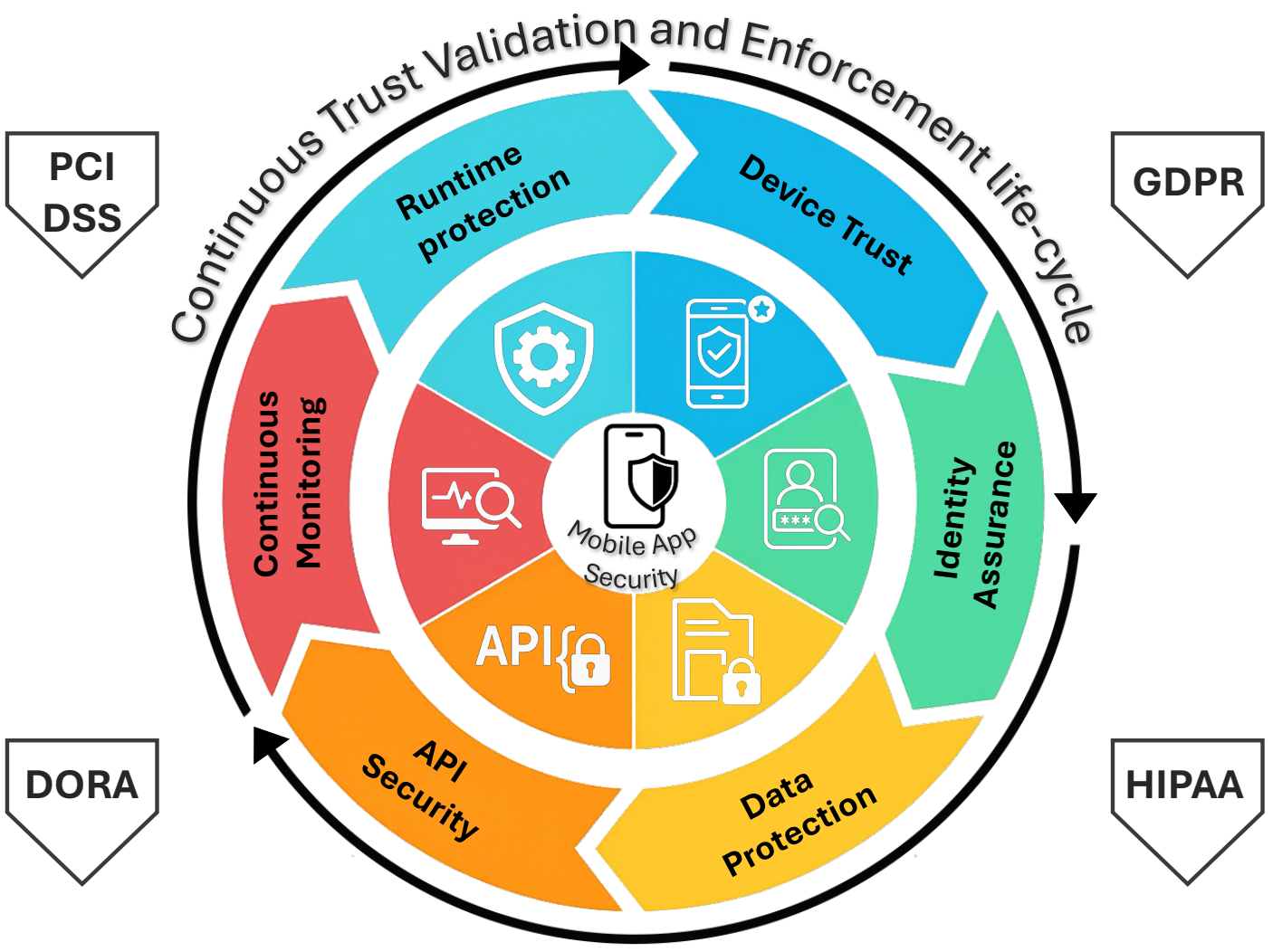

**Figure 2: Zero Trust Mobile Application Security Framework**

This contribution is intended as both a strategic reference and a practical guide. For academic audiences, it extends the conceptual boundaries of Zero Trust to encompass mobile-first realities. For practitioners—including CISOs, enterprise architects, and mobile security leaders—it offers a deployable roadmap to mitigate real-time threats, accelerate compliance, and strengthen digital trust. The framework addresses the urgency of the current threat landscape and provides an actionable model for institutions seeking to secure the mobile perimeter.

# 2.0 Literature and Industry Review

## 2.1 Foundations of Zero Trust Architecture

Zero Trust Architecture (ZTA) is built on continuous verification, least privilege, and context-aware access - principles formalized in NIST SP 800-207 and CISA's Zero Trust

Maturity Model (ZMM). These standards address dynamic policy enforcement, least-privilege access, device posture assessment, and application security. NIST SP 800-207 explicitly recognises personally owned devices within enterprise access scenarios, while CISA's Zero Trust Maturity Model includes applications and workloads executing on mobile devices (NIST, 2020a; CISA, 2023). More recent NIST implementation guidance provides detailed enterprise Zero Trust deployment examples, including support for hybrid workforce access (NIST, 2025). Although these publications provide important architectural and operational foundations, the specific challenge addressed in this paper is how to translate their principles into an integrated runtime assurance and enforcement model for consumer-facing mobile applications. Forrester's definition of modern Zero Trust provides broad strategic coverage but similarly lacks operational specificity for mobile ecosystems.

**Contrast with Recent Guidance:**
While these foundational documents establish the theoretical and operational basis of ZTA, they do not require organisational ownership of every endpoint. NIST SP 800-207 explicitly recognises unmanaged devices and distinguishes enterprise resource access from anonymous public and consumer-facing business processes, while acknowledging that selected Zero Trust policies may be applied to registered customers (NIST, 2020a). The remaining challenge is therefore not the applicability of Zero Trust principles to unmanaged devices, but the absence of an integrated, consumer-mobile-specific implementation model connecting runtime integrity, device posture, contextual authentication, and backend access enforcement.

## 2.2 Emerging Academic Perspectives and Adjacent Standards

Systematic reviews (Gambo & Almulhem, 2025) and research on Zero Trust in cloud and IoT (Liu et al., 2024) recognize the challenges of decentralization and the need for continuous authentication. However, these works highlight the absence of comprehensive runtime enforcement within the consumer mobile domain. Zohaib et al. (2024) and Azad et al. (2024) survey ZTNA and VPN but rarely address code integrity or runtime threats specific to mobile apps

**Expanded Academic Context:**
Li (2024) provides a detailed analysis of mobile application software security, highlighting that industry efforts are often skewed towards pre-deployment measures, leaving runtime exposures inadequately addressed. Ejiofor et al. (2025) examine the transition from perimeter-based network security towards Zero Trust Architecture, emphasising continuous verification, least-privilege access, adaptive access policies, and ongoing security monitoring. Although their analysis highlights the importance of dynamic trust evaluation across enterprise environments, it does not provide detailed implementation guidance for validating application integrity or enforcing contextual trust decisions within consumer-facing mobile application runtimes. Yenugula et al. (2023) explore the integration of Zero Trust Architecture and federated learning to prevent data leakage on enterprise mobile terminals. Their approach combines runtime-level mechanisms, including Android framework hooking, with transparent encryption, data isolation, and virtual remote desktop technologies. Although the study demonstrates an implementation focused on mobile data protection, it does not establish a comprehensive framework for continuously evaluating application integrity and enforcing contextual API access decisions throughout consumer-facing mobile application sessions.

**Debate Highlight:**
This academic discourse underscores a debate: Should mobile app security focus on static, pre-deployment hygiene or prioritize ongoing, in-session trust evaluation? The literature identifies a gap between regulatory intent and technical implementation, further complicated by the heterogeneity of mobile operating systems and threat surfaces.

## 2.3 Industry Reports and Threat Intelligence

Recent threat intelligence demonstrates that mobile security risks remain significant, although attack patterns vary across channels and threat categories. LexisNexis Risk Solutions (2026) reported a 56% year-over-year decline in the overall mobile app attack rate in 2025, alongside a 66% increase in the attack rate targeting password-reset interactions through mobile applications. Separately, Kaspersky (2025) documented a 196% increase in smartphone banking Trojan attacks during 2024, highlighting the continuing threat posed by mobile malware. Modern mobile apps integrate on average 15–30 SDKs (e.g., for analytics, payments, identity, and authentication), many of which operate with elevated privileges within the app context. These SDKs can become vehicles for exploitation if not independently protected. Threat modelling must account for SDK-specific attack vectors including binary lifting, re-signing, dynamic analysis, and misuse by hostile apps. Without runtime controls for embedded components, these third-party dependencies may undermine the trustworthiness of the overall mobile application. LexisNexis (2024) and IBM (2024) emphasize recent surges in data breaches tied to unprotected mobile app runtimes, supporting the argument that perimeter- or identity-centric controls are insufficient for stopping evolving mobile threats.

An additional and rapidly intensifying threat to mobile runtime security is the accelerated pace and accessibility of reverse engineering, driven by the convergence of automated decompiles and advanced large language models (LLMs). Tools such as Ghidra and JADX streamline binary lifting and static analysis, significantly reducing the technical threshold required to deobfuscate code or extract execution logic. Recent research demonstrates the growing potential of large language models (LLMs) to assist in understanding compiled binary code. Shang et al. (2024) evaluated LLMs across function-name recovery and binary-code summarisation tasks, demonstrating their ability to recover certain semantic information from binaries while identifying important limitations. Similarly, Tan et al. (2024) introduced LLM4Decompile, an open-source family of models designed to reconstruct and refine higher-level representations of compiled code, reporting improvements over conventional approaches on selected benchmarks. Together, these findings suggest that AI-assisted reverse engineering may reduce the effort required to analyse application logic, reinforcing the importance of layered code protection, runtime integrity verification, and complementary server-side security controls. However, these studies do not establish that arbitrary mobile binaries can be reliably reconstructed or that runtime protections can always be bypassed. As a result, runtime protections, anti-tampering measures, and API misuse defences must be designed to withstand not only conventional static analysis but also sustained inference attempts powered by AI-driven tools (Tan et al., 2024; Shang et al., 2024).

**Market Relevance:**
Despite forecasts that Zero Trust security market spending will exceed $78 billion by 2029 (MarketsandMarkets, 2024), recent industry studies show real-world mobile threat growth outpacing defensive investments — a persistent gap with significant cost implications (LexisNexis Risk Solutions, 2024; Kaspersky, 2025).

## 2.4 Regulatory and Technical Standards for Mobile Security

Regulatory frameworks like PSD2 RTS (European Commission, 2018), GDPR (EU Commission, 2024), and technical standards such as OWASP MASVS (2024), PCI-DSS (2024), and FIDO2 (FIDO Alliance, 2024) offer actionable controls for device validation, data protection, and authentication in mobile applications. PCI-DSS, for example, mandates strong authentication, session controls, and physical point-of-interaction device protection for payment applications, while the FIDO Alliance emphasizes scalable user authentication protocols better suited for diverse mobile endpoints.

**Comparative Analysis:**
Whereas enterprise-focused standards offer prescriptive device and network controls, mobile-centric standards still place a strong emphasis on user authentication and pre-deployment assessment, leaving runtime controls less developed. Emerging guidance from FIDO and ongoing updates to PCI-DSS reflect an evolution towards “continuous authentication,” but operational integration into real-time app sessions remains a challenge.

## 2.5 Recent Innovations and Proofs of Concept

Arikkat et al. (2025) introduce DroidTTP, a machine-learning-based approach that maps Android malware to tactics and techniques defined by the MITRE ATT&CK framework. By combining application feature analysis with classification models and large language models (LLMs), the study demonstrates how mobile threat intelligence can be enriched with information about adversarial behaviour. Although this approach offers valuable insights for threat detection and analysis, it does not establish continuous trust verification or enforcement within a live mobile application session. This distinction highlights the opportunity to integrate threat intelligence with runtime integrity assessment and context-aware access control in consumer-facing mobile applications.

## 2.6 Gap Analysis and Synthesis

Synthesizing across standards, academic research, and industry intelligence, it is clear that:

- **Gaps Remain in Consumer Mobile Runtime Operationalisation:** Existing Zero Trust architectures already address continuous verification, device posture, application access control, and security monitoring. NIST SP 800-207 explicitly considers personally owned devices, while CISA's Zero Trust Maturity Model incorporates application threat protection, contextual authorisation, and workloads executing on mobile devices (NIST, 2020a; CISA, 2023). However, these publications do not prescribe an integrated consumer-facing mobile application architecture that connects runtime integrity assessment, application-instance assurance, device risk signals, and backend policy enforcement throughout sensitive user interactions. This distinction is important because consumer applications operate on user-controlled devices where the service provider cannot assume administrative control or complete visibility into the execution environment. The research gap addressed in this paper is therefore one of operational integration and implementation specificity, rather than an absence of Zero Trust principles applicable to applications or unmanaged endpoints. The proposed six-pillar framework seeks to address this gap by coordinating application-level protection, contextual trust evaluation, and continuous monitoring within a coherent mobile security architecture.

- **Divergence in Practice:** There is inconsistency between standard-setting bodies' requirements (e.g., PCI-DSS and FIDO for payments/authentication) and actual Zero Trust deployment in the field (NIST, CISA).
- **Debates on Control Depth:** Reports from IBM and LexisNexis prompt ongoing debates about whether ZTA's primary investments should focus on network hardening or context-aware, in-app runtime enforcement.
- **Session-Start Trust versus Runtime Trust:** A growing body of practitioner evidence suggests that even solutions marketed as "Zero Trust" — including ZTNA agents and gateways — typically authenticate the user and device at connection establishment, then treat the resulting session as implicitly trusted until expiry or timeout. This pattern replicates the very assumption Zero Trust was designed to eliminate: that a verified starting condition guarantees ongoing trustworthiness. In consumer mobile environments, where device posture, network context, and runtime integrity can change mid-session, this limitation is particularly acute. The framework proposed in this paper addresses this by embedding enforcement within the application runtime itself, enabling per-operation trust evaluation rather than per-session trust assumption.
- **The Parallel Evolution of Mobile Security and the Persistent Zero Trust Gap:** Despite the growing adoption of Zero Trust Architecture (ZTA) in enterprise environments, its comprehensive application to mobile platforms has consistently lagged. This gap stems from a convergence of historical, technical, and perceptual factors. Enterprise security evolved within perimeter-based models, with ZTA emerging as a response to controlled, network-centric infrastructures (Broadcom, 2019; NIST, 2020a). In contrast, mobile security developed independently, focusing on device-level protection, app store vetting, and malware prevention (Samsung, 2019).

  In consumer-facing mobile environments, the challenge is not that Zero Trust principles depend on centrally managed devices, but that application providers have limited administrative control and visibility over customer-owned endpoints. This requires trust decisions to incorporate available application and device integrity evidence without assuming the management capabilities typically available in enterprise-controlled environments (NIST, 2020a).

Technically, platform fragmentation and runtime-specific threats—such as tampering, repackaging, and session hijacking—make consistent enforcement at the application layer challenging (ProAndroidDev, 2020; SpyCloud, 2024). Compounding this is a long-standing perception of mobile as a consumer-grade channel, leading to underinvestment, despite strong evidence of mobile's increasing role in fraud (LexisNexis Risk Solutions, 2024).

**Table 1: Key Contributions and Gaps in Prior Literature and Standards**

| **Reference/Standard** | **Mobile Runtime Controls** | **Focus Area** | **Noted Gaps/Contrasts** |
|---|---|---|---|
| NIST SP 800-207, CISA ZMM | General device, application and access-control guidance | Zero Trust architecture, contextual access and maturity | No integrated consumer mobile runtime-to-backend enforcement model prescribed |
| Li (2024) | Partial | Mobile app security | Emphasizes pre-deployment, not live controls |
| PCI-DSS, FIDO Alliance | Strong (auth/payment) | Real-time auth/compliance | Limited to identity/auth; light on runtime |
| Kaspersky, Experian, and IBM reports | Highlight gaps | Threat/fraud trends | Show rising fraud despite stronger policy base |
| Arikkat et al. (2025) | Not directly addressed | Android malware-to-TTP mapping using machine learning | Does not evaluate continuous runtime trust enforcement or contextual access decisions |
| Ejiofor et al. (2025) | Conceptual | Enterprise Zero Trust principles, continuous | Does not prescribe consumer mobile runtime integrity |
| | | verification and access governance | validation or in-app trust enforcement |

### 2.7 Summary

The literature collectively affirms the urgency of an operational Zero Trust model explicitly designed for the mobile runtime. Existing standards and research establish many of the relevant security principles and individual controls, but the sources reviewed here do not present a unified implementation model specifically connecting consumer mobile application runtime assurance with contextual, backend-enforced trust decisions. This paper's contribution addresses these articulated gaps by building a synthesized, standards-aligned, and threat-informed framework for Zero Trust in consumer mobile environments.

## 3.0 Methodology

Building on the identified gaps in current Zero Trust models, this section outlines the methodological approach and introduces a mobile-specific, six-pillar Zero Trust framework to address practical runtime threats.

This research applies a design science methodology to construct a novel framework for extending Zero Trust Architecture (ZTA) to mobile runtime environments. Design science is widely used in cybersecurity and systems engineering to develop innovative, actionable solutions grounded in both theory and empirical rigor.

### 3.1 Research Process

The framework was developed through a **three-stage methodology**:

a. **Analytical Phase:**
    - Evaluated existing Zero Trust models (e.g., NIST SP 800-207, CISA ZT Maturity Model) to identify core principles and gaps relevant to mobile ecosystems.

b. **Threat Landscape Synthesis:**

- Systematically reviewed threat intelligence from leading cybersecurity sources (OWASP MASVS, Kaspersky Mobile Threats Report, LexisNexis Risk Solutions Cybercrime Report) and recent academic findings.
- Classified runtime threats impacting consumer-facing mobile applications, such as malware, session hijacking, and API abuse.

c. **Practical Adaptation and Validation:**
- Mapped standards and regulatory mandates (e.g., OWASP MASVS, PSD2 RTS, GDPR) to operational mobility controls.
- Conducted iterative refinement via case studies and practitioner feedback to enhance real-world applicability and alignment with compliance requirements.

## 3.2 Data Sources

- **Primary Sources**: International cybersecurity standards, regulatory documents, threat intelligence reports, and peer-reviewed literature.
- **Secondary Sources**: Industry whitepapers, real-world case studies, and academic reviews of ZTA implementations.

## 3.3 Justification

This approach ensures:

- The framework is grounded in **established cybersecurity theory**, incorporates the **current threat landscape**, and remains **actionable** for practitioners.
- Alignment with **benchmark methodologies** established in academic cybersecurity research (e.g., systematic literature reviews, design-based frameworks)

The resulting model proposes six interdependent pillars of mobile-specific Zero Trust enforcement, each mapped to relevant control standards and designed for runtime integration.

# 4.0 Framework Design

## 4.1 The Six Pillars of Zero Trust for Mobile Applications

The proposed framework introduces six foundational control domains tailored for mobile runtime environments. These pillars represent the core of trust enforcement when enterprise control ends and user interaction begins.

This summary enables readers to anticipate the core focus of each pillar before examining their detailed implementation and standards alignment in the following subsections. To provide clarity before detailed examination, Table 2 presents an overview of each pillar's primary focus, enabling readers to understand the framework's comprehensive scope.

**Table 2: Overview of the Six Pillars of Zero Trust for Mobile Applications**

| **Pillar** | **Key Focus** |
|---|---|
| Runtime Protection | Prevent tampering, repackaging, and code manipulation |
| Device Trust | Verify device integrity and security posture |
| Identity Assurance | Continuously validate user and session authenticity |
| Data Protection | Safeguard data in transit, at rest, and during use |
| API Security | Enforce context-aware access and safeguard API traffic |
| Continuous Monitoring | Detect threats in real-time and trigger swift responses |

Each of these pillars addresses a distinct dimension of real-time mobile security and trust enforcement. The following subsections examine each pillar in detail, outlining its practical objectives, the threats it mitigates, and its alignment with relevant industry standards and regulatory frameworks.

**Pillar 1: Runtime Protection**

Within this framework, ‘Runtime Protection’ refers to the combined enforcement of code integrity, memory safety, and asset confidentiality - including static assets stored within the app package or sandbox, which are decrypted, validated, or protected at runtime. Security is embedded into the app itself. Self-defending capabilities such as anti-tampering, anti-debugging, and dynamic runtime integrity verification are applied. The runtime environment must not only secure the host application but also embedded components such as third-party SDKs. These SDKs often handle sensitive operations—payments, biometrics, or cryptography—and are therefore high-value targets for reverse engineering, tampering, and misuse. Effective runtime protection must ensure that these components exhibit the same integrity, obfuscation, and anti-tampering properties as the core application. These controls contribute to defence-in-depth by improving application resilience against tampering and runtime manipulation (OWASP, 2024). In regulated payment scenarios, selected integrity and anti-tampering measures may also support the safeguards described in Article 9(3) of the PSD2 RTS, particularly where strong customer authentication relies on a multipurpose device (European Commission, 2018).

**Objective:** Harden the mobile runtime against active manipulation and exploit.

The table below maps runtime security capabilities to OWASP MASVS v2.1.0 to illustrate operational alignment with the proposed Zero Trust enforcement layer.

**Table 3:** ***Pillar 1 - Illustrative Runtime Protection Capabilities Mapped to OWASP MASVS v2.1.0***

| Capability | Description | MASVS v2.1.0 |
|---|---|---|
| **Root/Jailbreak Detection** | Detects rooted or jailbroken devices to prevent execution in compromised OS environments. | MASVS-RESILIENCE-1 (Platform Integrity) |
| **Debugger/Hooking Detection** | Detects active debuggers or frameworks like Frida/Xposed used for runtime manipulation. | MASVS-RESILIENCE-4 (Anti-Dynamic Analysis) |
| **Emulator Detection** | Blocks execution on virtual devices used for automated analysis or evasion. | MASVS-RESILIENCE-4 (Anti-Dynamic Analysis) |
| **Code Injection Prevention** | Prevents memory modification or runtime injection of foreign code/libraries. | MASVS-RESILIENCE-2 (Anti-Tampering) |
| **Post-Compilation Obfuscation** | Obfuscates compiled binary to reduce code legibility and static analysis risk. | MASVS-RESILIENCE-3 (Anti-Static Analysis) |
| **Control Flow Obfuscation** | Disrupts predictable instruction flow, hindering reverse engineering. | MASVS-RESILIENCE-3 (Anti-Static Analysis) |
| **Checksum/Integrity Validation** | Validates the integrity of binaries and memory regions at runtime to detect tampering. | MASVS-RESILIENCE-2 (Anti-Tampering) |
| **Runtime Asset Protection** | Controls access to secrets or configuration in memory (not just on disk). | MASVS-STORAGE-1, MASVS-CRYPTO-1 |
| **SDK/Library Hardening** | Assess embedded third-party components for known vulnerabilities and apply appropriate integrity, obfuscation and anti-tampering measures where justified by the threat model. | MASVS-CODE-3 (components without known vulnerabilities) |
| **Repackaging & Signature Checks** | Detects repackaged/fake apps by verifying signing certificate or build metadata. | MASVS-RESILIENCE-2 (Anti-Tampering) |

**Pillar 2: Device Trust**

Devices are continuously assessed for security posture in real time, using signals such as rooting or jailbreaking status, emulator use, privilege escalation, and environmental tampering. In line with OWASP MASVS v2.1.0, particularly MASVS-RESILIENCE-1 (Platform Integrity), device posture assessment can incorporate signals associated with rooting, jailbreaking, emulation, and other environmental modifications. These signals support risk-informed access decisions but cannot guarantee that a device or application runtime is uncompromised. Contextual access decisions are then enforced based on the device's integrity and risk posture (PCI-DSS, 2024).

**Table 4: Alignment of the Device Trust Pillar with OWASP MASVS v2.1.0 Requirements**

| Device Trust Pillar | OWASP MASVS v2.1.0 Mapping | Key Protection Goal |
| --- | --- | --- |
| Real-time assessment of device posture (rooting, jailbreaking, emulator use, privilege escalation, environmental tampering) | MASVS-RESILIENCE-1: The app validates the integrity of the platform | Ensure applications execute only in trusted environments by detecting compromised devices |
| Contextual access enforcement based on device integrity | MASVS-PLATFORM-3: The app uses the user interface securely | Use platform integrity and device-risk evidence to inform policy-driven decisions for sensitive operations |
| Runtime validation of platform security features | MASVS-RESILIENCE-1: Platform integrity validation ensures OS security features can be trusted | Validate that platform security mechanisms (secure storage, biometrics, sandboxing) remain intact |

**Objective**: Block compromised devices from initiating high-risk transactions.

**Pillar 3: Identity Assurance**

This pillar enforces continuous authentication beyond the login event. Identity is verified dynamically throughout the session using secure elements, biometric binding, and cryptographic credential validation aligned with FIDO2 and NIST SP 800-63B (NIST, 2020b; FIDO Alliance, 2024).

**Objective:** Prevent session hijacking and impersonation.

**Pillar 4: Data Protection**

Data must be secured at rest, in transit, and in use – particularly during runtime, where traditional encryption methods may be insufficient. The framework applies identity-bound key encryption, secure enclaves, and in-app obfuscation based on NIST SP 800-57 (NIST, 2020c) and GDPR Article 32 (EU Commission, 2024).

To ensure comprehensive coverage, the model also incorporates secure local storage mechanisms for protecting sensitive data such as cryptographic keys, authentication tokens, and session artifacts. These mechanisms use software- or hardware-enforced encryption and enforce data access controls, even when devices are compromised (e.g., rooted or jailbroken).

As data traverses multiple components within a mobile application—including integrated third-party SDKs—it is critical that all execution units enforce consistent protection guarantees. This includes runtime encryption, obfuscation of data-handling logic, and checksum validation of SDK binaries to minimize the attack surface and reduce unauthorized access or leakage.

**Objective**: Safeguard sensitive data from leakage, interception, or manipulation.

**Pillar 5: API Security**

API interactions are treated as critical trust events. The app instance, session context, and device posture must be validated with every request. API token reuse, replay attacks, and data scraping are mitigated using mTLS and behavioural rate limiting (OWASP, 2024; EBA, 2024). This pillar aligns with the OWASP Mobile Application Security Verification Standard (MASVS) V2.1.0, specifically the MASVS-NETWORK category ("Network Communication Requirements"). MASVS-NETWORK defines controls to ensure that all network data is transmitted securely, protected against eavesdropping, tampering, and man-in-the-middle (MITM) attacks.

To further enforce Zero Trust principles, **API access must also validate the trustworthiness of the calling application instance**. This is achieved through **application-level attestation** mechanisms that cryptographically prove the integrity of the app and its runtime environment. These mechanisms ensure that API requests originate from:

- An unmodified, approved app binary

- A secure runtime (free from instrumentation, hooking, rooting, or jailbreaking)
- A trusted device profile

App attestation can be implemented using platform-native attestation services (e.g., hardware-backed verification), or through sovereign, cross-platform approaches that operate independently of OS vendors. Crucially, **attestation shifts API access control from identity-only to identity-plus-integrity**, aligning with Zero Trust's mandate of continuous verification.

OWASP MASVS-NETWORK-1 establishes secure network communication requirements, while MASVS-NETWORK-2 addresses identity pinning for remote endpoints under the developer's control. These controls strengthen transport security but do not independently establish application-instance integrity, prevent all replay attacks, or provide contextual API authorisation. The proposed framework therefore supplements MASVS-aligned transport protections with application integrity evidence and backend policy enforcement.

**Table 5: API Security alignment for OWASP**

| API Security Pillar | OWASP MASVS Mapping | Key Protection Goal |
|---|---|---|
| API Security | MASVS-NETWORK-1 and MASVS-NETWORK-2 | Secure transport, prevent MITM, enforce pinning, ensure confidentiality & integrity |
| App attestation | MASVS-RESILIENCE-1<br><br>MASVS-AUTH-1 | Use verifiable application and device evidence to inform risk-based backend access decisions without assuming absolute runtime integrity |

**Objective**: Ensure API requests originate only from authorised, untampered app sessions, validated through both transport-layer security and runtime attestation of app integrity.

**Pillar 6: Continuous Monitoring**

Static assessments are replaced by real-time behavioural and environmental monitoring. Telemetry feeds from mobile apps are integrated with enterprise SIEMs or SOAR platforms, enabling mid-session anomaly detection and policy enforcement (CISA, 2023; IBM, 2024).

**Objective:** Detect and respond to threats during live user sessions.

Figure 3 illustrates the six interdependent pillars of Zero Trust enforcement in mobile applications. Trust is established and continuously validated across the runtime lifecycle, beginning with app integrity and device posture, then advancing through identity assurance, data protection, contextual API enforcement, and continuous observability.

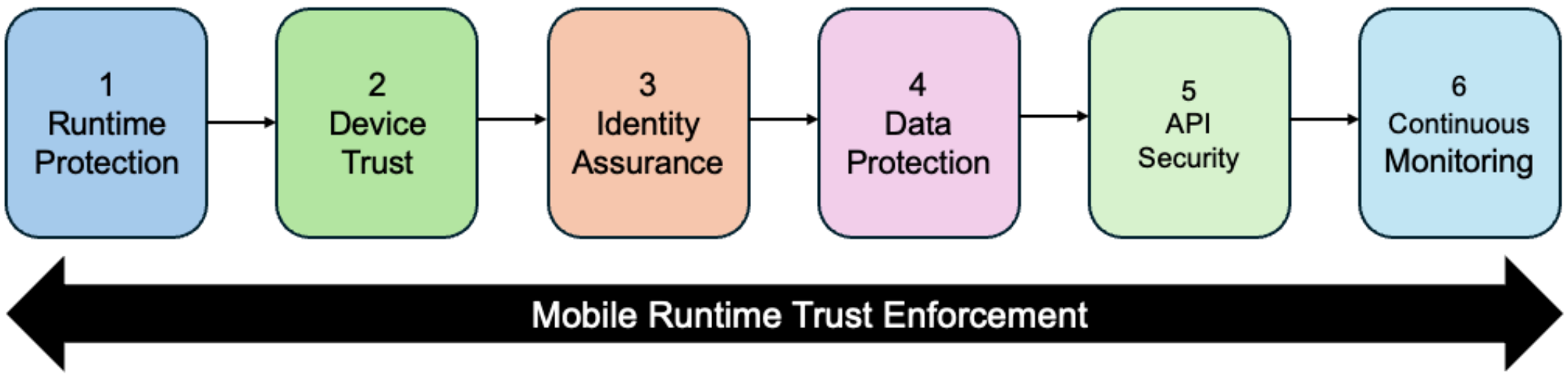

**Figure 3: Sequential Pillars of Mobile Runtime Zero Trust Enforcement**

## 4.2 Standards Alignment and Control Mapping

Each pillar is explicitly mapped to established security and compliance standards, as presented in Table 6 below:

**Table 6: Mapping of Zero Trust Mobile Pillars to Security Standards and Regulatory Frameworks**

| Zero Trust Pillar | NIST SP 800-207 Core Principles & Components | OWASP MASVS Controls | PSD2 RTS Requirements | GDPR Articles | CISA ZTMM Domains |
|---|---|---|---|---|---|
| **Runtime Protection** | Runtime enforcement assuming breach; PEPs adapt policies dynamically | MASVS-RESILIENCE-1/2/3/4 | Article 9(2)–(3): Independence of SCA elements on multipurpose devices, including secure execution and software/device alteration safeguards | Art 32: Security of Processing | Applications & Workloads |

| Zero Trust Pillar | NIST SP 800-207 Core Principles & Components | OWASP MASVS Controls | PSD2 RTS Requirements | GDPR Articles | CISA ZTMM Domains |
|---|---|---|---|---|---|
| **Device Trust** | Trust evaluation based on device posture and signals | MASVS-RESILIENCE-1 | Article 9(3): Software/device integrity safeguards supporting the independence of SCA elements; not a general device-trust mandate | * Articles 5(1)(f) and 32 | Devices |
| **Identity Assurance** | Continuous identity verification; dynamic PDP context evaluation | MASVS-AUTH-1/2/3 | Articles 4–9: Authentication code, dynamic linking, authentication factors and independence of SCA elements | Art 32: Security of Processing | Identity |
| **Data Protection** | Least privilege and per-session access control | MASVS-STORAGE-1/2; MASVS-CRYPTO-1/2 | Article 22: Confidentiality and integrity of personalised security credentials; Article 35: Secure communications between relevant payment service providers | Art 25 & 32: Privacy by Design, Security of Processing | Data |
| **API Security** | Context-aware dynamic API access control via Policy Engine | MASVS-NETWORK-1/2; MASVS-AUTH-1 for relevant authentication and authorisation practices | Articles 30–35: Access interfaces, identification and secure communications between relevant payment service providers; not general mobile API authorisation | Art 32: Security Measures | Applications & Workloads |
| **Continuous Monitoring** | Telemetry collection by PEP/PE; analytics-driven decisions | ** MASVS-RESILIENCE: Anti-Tampering, Anti-Dynamic Analysis" paired with telemetry capabilities. | Article 2: Transaction monitoring to detect unauthorised or fraudulent payments; not a requirement for in-app SIEM telemetry | Art 33 & 34: Breach Notification & Communication | Visibility & Analytics |

* While GDPR does not explicitly require device trust, Articles 5(1)(f) and 32 imply the need to ensure device security as part of broader obligations to maintain data confidentiality and processing integrity.

** Note: MASVS v2.1.0 does not have a dedicated "telemetry" or "runtime monitoring" control. Telemetry integration is an operational extension beyond MASVS scope, mapped instead to CISA ZTMM Visibility & Analytics.

This mapping identifies how selected framework controls may contribute to relevant security and regulatory objectives while supporting integration with existing enterprise Zero Trust roadmaps. Actual regulatory compliance depends on the organisation's applicable obligations, implementation scope, operating effectiveness, and supporting evidence, and cannot be established through architectural control mapping alone.

Beyond the core standards mapped in **Table 6**, the framework has been explicitly aligned to three additional sector-critical regulatory standards: PCI DSS 4.0, HIPAA, and the EU Digital Operational Resilience Act (DORA). The full control-by-control mapping is presented in Appendix Table A1, and a regulatory-aligned maturity progression showing how each implementation phase addresses specific compliance requirements is provided in Appendix Table A2. These mappings are designed to support compliance reporting, procurement baseline documentation, and audit evidence preparation across finance, healthcare, and EU-regulated sectors.

## 4.3 Design Rationale and Differentiation

Rather than replacing established enterprise Zero Trust architectures, the proposed framework extends their principles of continuous verification, least privilege, and contextual access control into the execution environment of consumer-facing mobile applications. Its distinctive contribution is the integration of runtime integrity assessment, device posture, identity assurance, data protection, API security, and continuous monitoring into a coordinated model in which application-level trust signals inform policy-driven access decisions for sensitive mobile operations.

Furthermore, the modularity of the framework enables phased adoption, which will be explored in the next section. This approach supports organisations in aligning security investments with risk tolerance, compliance urgency, and application development maturity.

**Distinctive Advantages of the Six-Pillar Model:**

- **Comprehensive Depth:** Consolidates trust across device health, code integrity, authenticated identity, and real-time user behaviour—addressing the fragmentation seen in legacy solutions.

- **Contextual Breadth:** Operates directly within the mobile application's execution context, enforcing controls where modern threats typically manifest.
- **Standards Alignment:** Provides auditable mapping to OWASP MASVS, GDPR Article 32, and PSD2 RTS, enhancing regulatory readiness.
- **Complementarity with Existing Zero Trust Investments:** Operates at the application runtime layer, complementing rather than duplicating network and access-layer investments such as SASE and ZTNA. Where SASE governs the secure path between app and backend, this framework governs the integrity of the app endpoint generating that traffic.

**Conclusion and Implications:**

- Point solutions offer isolated control and fail to deliver unified runtime assurance.
- The proposed Zero Trust model embeds enforcement at the session level, enabling proactive, real-time defence across critical trust layers.
- Its phased adoption pathway and alignment with maturity metrics provide a practical route to reducing risk and achieving demonstrable compliance.

This comparative analysis underscores the operational value of a mobile-specific Zero Trust architecture. It not only bridges critical control gaps left by conventional solutions, but also elevates mobile security to a posture of continuous, verifiable trust enforcement - necessary for today's high-risk consumer application landscape.

Having established the framework's core components, the next section outlines a phased roadmap for practical deployment within diverse mobile environments.

# 5.0 Phased Implementation Roadmap

## 5.1 Overview

Having established the framework's theoretical foundation, attention now turns to practical deployment. Implementing Zero Trust Architecture (ZTA) in mobile environments requires an approach that is adaptive to operational constraints, device heterogeneity, and user behaviour. Unlike traditional enterprise systems, mobile applications operate in decentralised, adversarial conditions where runtime manipulation, environmental tampering, and spoofed signals can occur undetected if foundational controls are not prioritised.

This roadmap introduces a **three-phase deployment model** optimised for mobile-first use cases. Each phase is structured around a 30-day implementation horizon, with controls sequenced according to risk mitigation priority and interdependency. The model emphasises **runtime assurance first**, before layering in identity, data, and behavioural trust signals. This ordering aligns with real-world threat vectors, OWASP MASVS resilience controls (OWASP, 2024), and regulatory standards, including PSD2 RTS and GDPR.

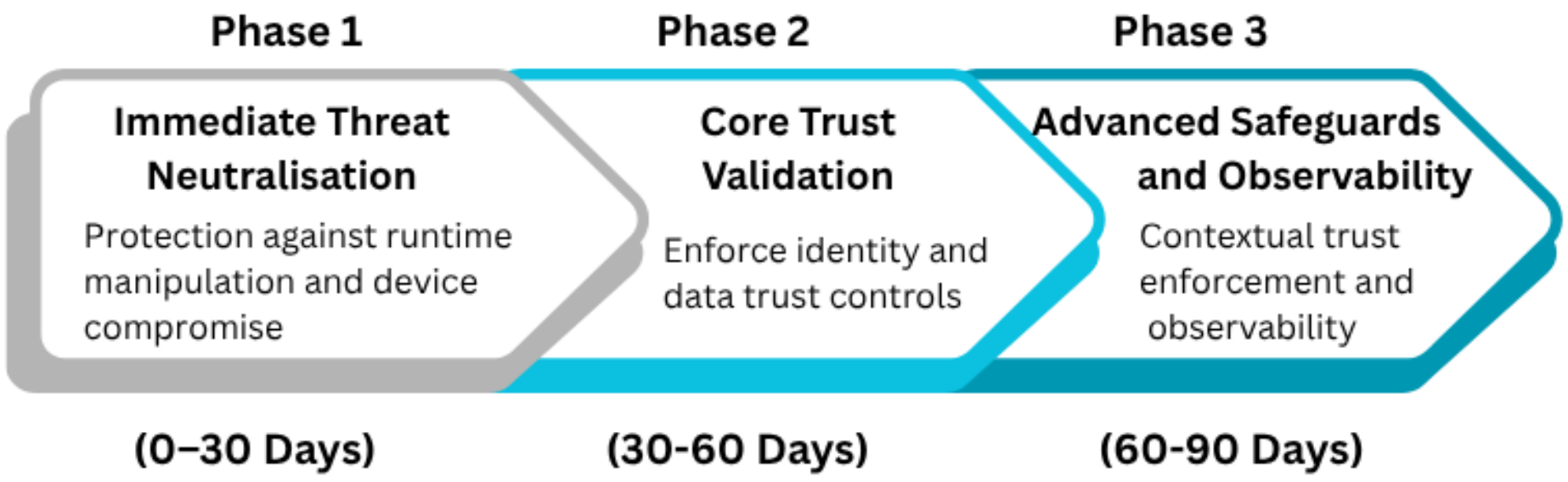

**Figure 4: Phased Implementation Roadmap for Zero Trust in Mobile Applications**

The visual timeline in Figure 4 illustrates the three-phase implementation strategy for operationalising Zero Trust in mobile environments. Each phase builds upon the previous one, beginning with runtime and device-level hardening (Phase 1), advancing through identity and data trust enforcement (Phase 2), and culminating in API-level security and real-time monitoring (Phase 3). The design reflects a logical, risk-prioritised progression, aligning with regulatory mandates and mobile-specific threat vectors.

## 5.2 Phase 1: Immediate Threat Neutralisation (0–30 Days)

This phase establishes baseline trust by mitigating the most urgent threats — runtime manipulation and device compromise. By addressing the app’s execution context first, organisations ensure that all subsequent controls (e.g., authentication, API requests) operate within a validated, tamper-resistant environment.

Key Capabilities:

- **Runtime Protection:** Implement anti-tampering, anti-debugging, and code obfuscation mechanisms (MASVS-RESILIENCE).
- **Device Trust Assessment:** Enforce access denial for rooted, jailbroken, emulated, or compromised devices (MASVS-RESILIENCE-1).
- **Secure Bootstrapping:** Integrate posture checks at app launch, prior to identity prompts or sensitive data access.

**Rationale:**
Zero Trust begins with the assumption of compromise. In mobile, this compromise is most likely at the runtime layer. Identity verification conducted within a compromised mobile environment may be vulnerable to credential interception, manipulation, or other forms of interference, reinforcing the need for additional device and runtime integrity controls (Kaspersky, 2023).

Critically, runtime protection is typically deployed via SDK integration into existing application build pipelines rather than requiring full application rewrite, significantly lowering the adoption barrier and enabling organisations to achieve Phase 1 maturity within the proposed 30-day horizon even for production applications already in market.

**Outcomes:**

- Known mobile malware vectors are mitigated before trust is granted
- Only untampered app instances may proceed to sensitive workflows
- Foundational telemetry is collected for continuous trust evaluation

## 5.3 Phase 2: Core Trust Validation (30–60 Days)

With a hardened runtime baseline, organisations can now enforce identity and data trust controls. This phase strengthens authentication, binds session identity to device context, and ensures secure data handling aligned with regulatory requirements.

**Key Capabilities:**

- Identity Assurance: Deploy strong customer authentication (SCA) with biometric or cryptographic binding (FIDO Alliance, 2024; NIST, 2020b).
- Session Trust Binding: Use PKCE (Proof Key for Code Exchange) in OAuth flows to bind authentication tokens to runtime posture and device context, preventing reuse or replay. Follow OAuth 2.1 guidelines for secure token lifecycle management.
- Data Protection: Apply identity-bound encryption for sensitive data at rest, secured via secure enclaves or platform-specific key stores (NIST, 2020c; EU Commission, 2024).
- **Runtime Component Trust:** Extend runtime trust enforcement to all embedded components within the application, including third-party SDKs and native libraries. These components often handle sensitive logic and operate with elevated privileges, yet they are frequently overlooked in security strategies. Trust boundaries must be applied uniformly across the entire app execution boundary, ensuring that any embedded code is subject to the same runtime integrity, anti-tampering, and obfuscation controls as the host application.

**Rationale**:
Runtime integrity assessment should complement existing authentication controls rather than delay their implementation. Where strong customer authentication relies on a multipurpose device, Article 9 of the PSD2 RTS establishes requirements concerning the independence of authentication elements and safeguards against software or device alteration. These requirements support the inclusion of runtime integrity controls within a defence-in-depth strategy but do not prescribe a universal implementation sequence in which authentication follows runtime hardening (European Commission, 2018). Extending trust enforcement uniformly to embedded components mitigates risks posed by privileged third-party code, which can become vectors for exploitation if unprotected.

**Outcomes**:

- Users are continuously re-evaluated during sessions based on device and embedded component behaviour.
- Sensitive data is cryptographically tied to session integrity.
- Embedded SDKs and native libraries maintain enforced runtime integrity, reducing the risk of tampering or misuse.
- Regulatory authentication and encryption mandates are satisfied.

## 5.4 Phase 3: Advanced Safeguards and Observability (60–90 Days)

The final phase enables dynamic, contextual trust enforcement and observability. Runtime signals, behavioural telemetry, and contextual access policies are operationalised to defend against sophisticated threats such as session hijacking, credential stuffing, or behavioural fraud.

**Key Capabilities:**

- **API Security Enforcement:** Apply context-aware validation of sensitive API requests using application integrity evidence, device posture, session risk, and runtime signals. This represents a proposed architectural extension of Zero Trust principles, complementing secure transport requirements in OWASP MASVS-NETWORK and relevant authentication and transaction monitoring provisions of the PSD2 RTS (OWASP, 2024; European Commission, 2018).
- **Continuous Monitoring:** Feed mobile telemetry to enterprise SIEM/SOAR platforms to detect anomalies and automate policy response (CISA, 2023). Telemetry schemas should align with standards such as OpenTelemetry to enable consistent parsing, enrichment, and correlation across enterprise observability platforms.
  • **Adaptive Response:** Implement in-app risk scoring, rate limiting, and real-time trust revocation mechanisms.

**Rationale:**
API abuse and behavioural fraud are among the fastest-growing attack vectors (LexisNexis Risk Solutions, 2024). This phase shifts from static controls to intelligent, behaviour-informed trust decisions. Standardised telemetry enhances interoperability, making Zero Trust enforcement auditable and resilient. It completes the Zero Trust model by enabling feedback-driven access governance.

**Outcomes**:

- Suspicious behaviour triggers live session adaptation or revocation
- API abuse is detected and throttled dynamically
- Mobile telemetry becomes part of enterprise-wide risk visibility

## 5.5 Governance and Iteration

This roadmap is designed to support modular implementation based on application criticality, compliance obligations, and business risk appetite. It should be governed by a cross-functional working group including mobile engineering, cybersecurity, fraud prevention, and compliance leaders. Periodic reassessment of threat models and telemetry feedback should inform adjustments to control sequencing and policy thresholds.

By emphasising early runtime hardening and sequenced identity enforcement, this model offers a pragmatic, risk-aligned path toward operationalising Zero Trust principles in high-risk mobile environments.

To support practical adoption and benchmarking, the following matrix maps each pillar of the Zero Trust for Mobile framework to a three-level maturity model aligned with the phased roadmap. It enables organisations to assess their current state, prioritise next steps, and measure progress over time. The matrix shown in Table 5 outlines progressive levels of control implementation for each pillar, aligned to the Immediate Threat Neutralisation, Core Trust Validation, and Advanced Safeguards phases. Each level defines measurable security outcomes across runtime integrity, trust enforcement, and observability.

**Table 7: Maturity Assessment Matrix: Zero Trust for Mobile Applications**

| Zero Trust Pillar | Phase 1: Immediate Threat Neutralisation *(0–30 Days)* | Phase 2: Core Trust Validation *(30–60 Days)* | Phase 3: Advanced Safeguards & Observability *(60–90 Days)* |
|---|---|---|---|
| Runtime Protection | App hardening via anti-tampering and basic obfuscation | Runtime integrity validation during session lifecycle | Behaviour-triggered self-defence (e.g., dynamic RASP) |
| Device Trust | Root/jailbreak/emulator detection with access enforcement | Risk-based device scoring and session correlation | Continuous posture reassessment with policy-driven action |
| Identity Assurance | — (not yet enforced) | Biometric/FIDO2-based authentication tied to device context | Adaptive identity validation with behaviour-based re-authentication |
| Data Protection | Static encryption of sensitive data at rest | Identity-bound keys, secure enclave usage | Context-aware access revocation and encryption key rotation |
| API Security | — (minimal enforcement) | Session-based API authentication with contextual gating | Real-time API request scoring and dynamic throttling |
| Continuous Monitoring | — (no telemetry integration) | Basic telemetry collected and logged centrally | Runtime telemetry integrated with SIEM/SOAR for auto-response |

This structured maturity progression can serve as a diagnostic tool for architecture teams, as well as a planning asset for CISO dashboards, Zero Trust program boards, or compliance reporting.

The regulatory-aligned maturity progression in Appendix Table A2 demonstrates how Phase 1 controls address PCI DSS secure development and access requirements (6.2.3, 8.4.2) and Phase 2 aligns with HIPAA, while Phase 3 fulfils DORA's detection and incident management requirements (Art. 10, Art. 17).

The following visual heatmap illustrates the progressive maturity levels of each Zero Trust pillar, aligned to the three implementation phases. The gradient highlights the evolution from foundational hardening to adaptive safeguards and runtime intelligence, enabling organisations to benchmark readiness and prioritise investments.

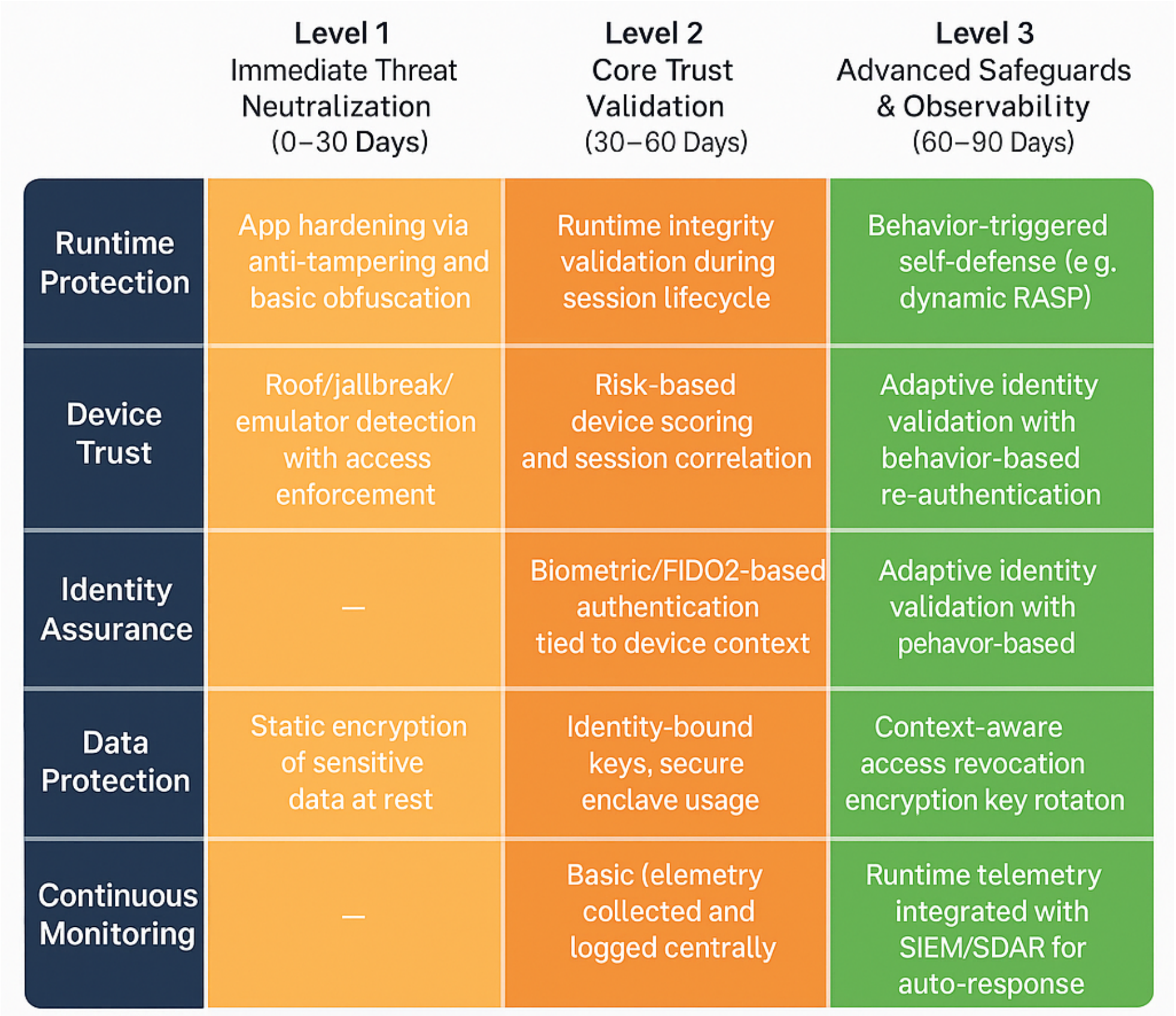

| | Level 1 Immediate Threat Neutralization (0–30 Days) | Level 2 Core Trust Validation (30–60 Days) | Level 3 Advanced Safeguards & Observability (60–90 Days) |
|---|---|---|---|
| Runtime Protection | App hardening via anti-tampering and basic obfuscation | Runtime integrity validation during session lifecycle | Behavior-triggered self-defense (e g. dynamic RASP) |
| Device Trust | Roof/jallbreak/ emulator detection with access enforcement | Risk-based device scoring and session correlation | Adaptive identity validation with behavior-based re-authentication |
| Identity Assurance | — | Biometric/FIDO2-based authentication tied to device context | Adaptive identity validation with pehavor-based |
| Data Protection | Static encryption of sensitive data at rest | Identity-bound keys, secure enclave usage | Context-aware access revocation encryption key rotaton |
| Continuous Monitoring | — | Basic (elemetry collected and logged centrally | Runtime telemetry integrated with SIEM/SDAR for auto-response |

**Figure 5: Heatmap of Maturity Levels Across Zero Trust Pillars for Mobile Applications**

Having established the framework's core components, the next section provides a phased deployment roadmap to guide organisations in sequencing and operationalizing these control domains within live mobile environments.

# 6.0 Zero Trust Mobile KPI Framework

## 6.1 Integrating KPIs and Maturity Metrics into Mobile Zero Trust

To operationalise the Zero Trust model in mobile application environments, it is crucial not only to implement technical controls but also to measure their effectiveness systematically. The following **Key Performance Indicators (KPIs)** and **Maturity Model** are designed to assist organisations in tracking Zero Trust adoption across three implementation phases. These instruments are informed by established frameworks, including OWASP MASVS, CISA Zero Trust Maturity Model, NIST SP 800-207, and Verizon DBIR.

KPIs should be monitored regularly (e.g., weekly), integrated into DevSecOps workflows, and visualised through dashboards. Meanwhile, the maturity model serves as a diagnostic and planning tool to identify gaps, benchmark progress, and guide investment priorities in line with Zero Trust objectives.

**Table 8: KPI Mapping Across Zero Trust Maturity Phases for Mobile Security**

| Phase | Capability | KPI | Measurement Frequency | Reference |
|---|---|---|---|---|
| **Phase 1: Immediate Threat Neutralisation** | Runtime Protection | % of apps with RASP enabled in production | Weekly | MASVS-RESILIENCE MAS-SIG 2023 |
| | Device Trust Assessment | % of sessions denied for rooted/emulated devices | Weekly | NIST IR 8144, MASVS-RESILIENCE-1 |
| | Secure Bootstrapping | % of app launches with posture check pre-authentication | Weekly | MASVS-RESILIENCE-1, NIST SP 800-207 |
| **Phase 2: Core Trust Validation** | Identity Assurance | % of SCA flows using biometrics/FIDO2 | Weekly | NIST SP 800-63B, PSD2 RTS Articles 4–9 |
| | Session Trust Binding | % of OAuth flows using PKCE & state param | Weekly | OAuth 2.1 Draft, MASVS-NETWORK |
| | Data Protection | % of sensitive data encrypted with session-tied keys | Weekly | GDPR Article 32, NIST SP 800-57 |
| **Phase 3: Advanced Safeguards & Observability** | API Security | % of API calls blocked based on device/session risk | Weekly | OWASP API Top 10, MASVS-NETWORK |
| | Continuous Monitoring | % of telemetry events ingested into SIEM/SOAR | Weekly | CISA ZTMM, OpenTelemetry 2023 |
| | Adaptive Response | MTTD & MTTR for risky sessions | Weekly | PCI DSS v4.0, CISA Metrics Guide |

**Table 9: Zero Trust Implementation Maturity Levels for Mobile Applications**

| Level | Description | Sample KPIs |
|---|---|---|
| **Level 1: Reactive** | No runtime enforcement; mobile app defences depend on external controls | • No RASP in prod• Mobile fraud rate untracked• No device posture validation |
| **Level 2: Foundational** | Runtime protections and device trust mechanisms in place | • Tamper detection coverage• % of launches blocked for compromised devices |
| **Level 3: Controlled** | Identity, session, and data controls tied to runtime state | • % of flows using PKCE• Encryption tied to session tokens• Risk-based re-authentication events |
| **Level 4: Adaptive** | In-app context drives access decisions and trust scoring | • Real-time API enforcement• Live session throttling• Mobile SIEM telemetry integration |
| **Level 5: Integrated** | Mobile trust telemetry informs enterprise risk engines and policy | • Cross-platform access rules tied to mobile trust• Detect-to-response time <15 mins• Measurable reduction in mobile fraud |

These KPIs and maturity benchmarks provide a structured mechanism for evaluating Zero Trust adoption within mobile environments. By aligning measurable indicators to each implementation phase and maturity level, security teams can ensure that progress is both strategic and demonstrably effective. As organisations advance along this model, they gain increased assurance that sensitive mobile workflows are protected, even in high-risk and unmanaged contexts, through continuous and context-aware enforcement rooted in verified runtime conditions.

# 7. Applied Architecture: Operationalising Zero Trust in Mobile Runtime

To demonstrate the practical application of the proposed six-pillar framework, this section explores its integration within a hypothetical mobile banking application. The scenario illustrates how Zero Trust principles are embedded across the mobile session lifecycle, emphasising enforcement at runtime rather than relying solely on perimeter controls. Each security mechanism corresponds to a specific pillar, ensuring comprehensive coverage aligned with current best practices.

## 7.1 Core Runtime Enforcement Mechanisms

### 1. Runtime-Aware Session Validation

Each user session begins with biometric authentication (e.g., FaceID) and secure enclave-backed device posture attestation. These ensure that both the user and the execution environment meet minimum assurance thresholds before access is granted. The control maps directly to Device Trust and Identity Assurance, as defined in Pillars 2 and 3 (NIST, 2020a; OWASP, 2024).

### 2. Contextual Encryption for Transaction Integrity

Sensitive operations, such as initiating a fund transfer or submitting a payment form, are protected using identity-scoped encryption keys. These keys are stored in platform-specific secure elements (e.g., iOS Keychain, Android Keystore) and are only accessible within untampered sessions. This protection directly supports Data Protection under Pillar 4 (NIST, 2020c; GDPR Art. 25 & 32).

### 3. Adaptive API Access Control

API endpoints are protected by policies that evaluate runtime signals such as device risk scores, geolocation, and app behaviour. Requests from degraded or suspicious runtime environments are denied or throttled. This proposed control operationalises API Security under Pillar 5 by combining secure network communications with contextual, backend-enforced authorisation. It complements OWASP MASVS-NETWORK and, where applicable, PSD2 RTS requirements concerning authentication, transaction monitoring, and secure communications, without treating those standards as mandating this specific runtime enforcement mechanism (OWASP, 2024; European Commission, 2018).

**4. Threat-Aware Observability and Response**

In-app telemetry, including runtime anomalies, behavioural deviations, and environmental flags, is streamed to an external SIEM. This enables real-time monitoring, threat hunting, and automated response workflows. These capabilities fulfil the Continuous Monitoring and Runtime Protection objectives of Pillars 1 and 6 (CISA, 2023; (LexisNexis Risk Solutions, 2024).

To further clarify the operational application of the six-pillar Zero Trust framework within real-world mobile environments, Figure 5 provides a visual summary of how these core principles are embedded throughout the lifecycle of a typical mobile banking session. This illustration demonstrates the practical integration of runtime, device, identity, data, API, and monitoring controls within the user journey, highlighting the sequential enforcement of trust and the interdependencies between security components.

The following section explores key implementation considerations and outlines areas where further research is needed to support effective, scalable adoption.

## 7.2 Implementation Considerations and Research Implications

This applied architecture demonstrates how Zero Trust principles can be embedded throughout the mobile runtime lifecycle. Translating this conceptual framework into a real-world deployment, however, raises critical implementation challenges and opens several avenues for further research.:

a. Balancing Security Depth and User Experience
    a. Understanding the impact of runtime enforcement on app latency, biometric flow interruptions, and battery consumption remains critical for optimising both protection and usability.

b. CI/CD Integration and Deployment Pipelines
    a. Investigating methods for embedding Zero Trust controls into mobile DevSecOps pipelines can enhance deployment velocity without compromising security baselines.
c. Telemetry Schema Standardisation
    a. Developing unified telemetry formats for in-app risk signals may improve vendor interoperability, traceability, and audit readiness across platforms and security stacks. Future research should evaluate common telemetry schemas, including their ability to support consistent risk assessment and integration with enterprise security monitoring systems.

### 7.3 Strategic Positioning

By treating the mobile application itself as an active enforcement node - rather than a passive interface - this architecture extends Zero Trust to the edge of user interaction. It closes the gap between conceptual Zero Trust models and operational requirements for mobile-centric organisations operating in high-risk or regulated sectors.

Emerging practitioner consensus reinforces this architectural direction. Industry discussions increasingly describe a convergence pattern at the runtime boundary — *attest, authorise, connect, observe, re-evaluate* — in which the enforcement point moves inside the application process itself, and the network becomes an implementation detail rather than the trust anchor (Griffiths, 2025). This convergence validates the framework's positioning of runtime protection as the foundational layer upon which identity, data, API, and monitoring controls are sequenced. Whether described as converging enforcement planes or as interdependent pillars, the direction of travel is consistent: from perimeter trust to runtime trust, with the live app session as the decisive enforcement boundary.

These practical insights and open questions set the stage for broader industry and academic engagement, as detailed in the concluding section.

## 8. Conclusion and Call for Collaboration

Having demonstrated the framework's operational relevance through a banking scenario, it is clear that closing the mobile trust gap requires both robust in-app security measures and a collaborative ecosystem approach.

The six-pillar reference model and associated maturity roadmap provide a deployable foundation for organisations to prevent fraud, enhance compliance, and deliver resilient mobile-first services. Yet the evolving threat landscape and diversity of mobile platforms demand continual validation and adaptive strategies.

**Sustained progress in this space will rely on open collaboration across research, enterprise, and standards bodies**. Future work should prioritise multi-sector field trials, open schema development for secure telemetry, and harmonization of assurance metrics. Only through shared effort can the community accelerate adoption and ensure trust in next-generation mobile digital experiences.

This paper encourages practitioners, researchers, and policymakers to collectively extend, test, and refine Zero Trust models for the mobile runtime. The journey toward ubiquitous runtime trust is just beginning—and requires collective leadership and innovation.

The increasing interdependence between mobile applications and high-assurance digital services, including payments, identity, and healthcare, exposes a critical trust gap in how organisations secure app runtimes. Although established Zero Trust architectures already address mobile devices, contextual access decisions, and application security, consumer-facing mobile applications present distinct operational challenges because they execute on user-controlled endpoints where runtime integrity, application authenticity, and device risk must be evaluated without relying on centralised device administration.

This paper proposes a six-pillar reference model for embedding Zero Trust directly into the mobile runtime environment. By prioritising runtime protection and device trust as foundational layers, and sequencing identity, data, API, and observability controls in a staged implementation roadmap, the framework reflects real-world threat dynamics and operational constraints. The maturity model and applied architecture scenario position this approach as both implementable and measurable.

As mobile attack vectors— including overlay fraud, credential stuffing, and advanced reverse engineering techniques — continue to evolve, organisations must shift from passive detection toward active enforcement of trust within the application itself. This paradigm requires new collaboration across security researchers, mobile engineers, policy experts, and platform stakeholders.

## 8.1 Areas for Future Collaboration and Research

**Empirical Validation:** Field-testing the roadmap and maturity model in multiple sectors such as finance, digital identity, and healthcare, with the goal of evaluating both control efficacy and operational feasibility.

**Cross-Industry Standardisation**: Defining shared schemas for runtime telemetry, trust signals, and security event reporting to support vendor-neutral implementations and regulatory auditability.

### 8.2 Final Reflection

In a mobile-first threat landscape, where applications increasingly serve as both the point of compromise and the last line of defence, Zero Trust must extend into the runtime itself. This framework provides a scalable blueprint designed to meet emerging regulatory demands while delivering practical controls aligned with evolving mobile threat dynamics.

This paper invites researchers, technology leaders, and public-sector stakeholders to collectively expand upon, validate, and accelerate the development of Zero Trust paradigms purpose-built for the modern mobile ecosystem. The journey toward ubiquitous runtime security is underway—success depends on shared leadership, continued innovation, and a commitment to operational excellence across sectors.

### 8.3 Key Takeaways:

- Zero Trust for mobile must operationalise continuous, in-app defence, not just perimeter or identity controls.
- The proposed six-pillar framework provides a standards-aligned, practical blueprint for organisations facing real-world mobile threats.
- Phased, maturity-driven implementation supports risk-based prioritisation and measurable progress.
- Cross-industry validation, open telemetry schemas, and collaboration will be essential to close the mobile trust gap.

# 10.0 APPENDIX

## 10.1 Extended Regulatory Mapping: PCI DSS 4.0, HIPAA, and DORA

To address sector-specific compliance requirements, the six-pillar Zero Trust Mobile framework is explicitly mapped to three critical regulatory standards:

1. **PCI DSS 4.0** (Payment Card Industry Data Security Standard)
2. **HIPAA** (Health Insurance Portability and Accountability Act Security Rule)
3. **DORA** (Digital Operational Resilience Act, EU Regulation 2022/2554)

This alignment enables organisations in finance, healthcare, and regulated sectors to operationalise Zero Trust while fulfilling compliance mandates.

**Table A1: Zero Trust Mobile Pillars Mapped to PCI DSS 4.0, HIPAA, and DORA**

| Zero Trust Pillar | PCI DSS 4.0 Requirements | HIPAA Security Rule Requirements | DORA Requirements |
|---|---|---|---|
| **Runtime Protection** | • **6.2.3** (code reviews for custom code),<br>• **6.2.4** (protect against common attacks),<br>• **6.5.1** (address vulnerabilities before production) | • §164.312(c)(1): Integrity controls | • Art. 9: (Protection and prevention) |
| **Device Trust** | PCI DSS 4.0 does not have a direct equivalent for mobile device posture assessment; | — | • Art. 9: (Protection and prevention — covers ICT security measures) |
| **Identity Assurance** | • **8.2.1** (unique user ID ✓),<br>• **8.3.6** (MFA for remote access),<br>• **7.2.1** (least privilege ✓) | • §164.312(a)(1): Unique user IDs<br>• **§164.312(a)(2)(iii)** (automatic logoff — addressable)<br>• §164.312(d): Entity authentication | • **Art. 9(4)(c)** (authentication mechanisms and access control policies) |
| **Data Protection** | • 3.5.1: Encryption of stored data | • §164.312(a)(2)(iv): ePHI encryption | • **Art. 9(4)(b)** (encryption and data protection), |
| | • 4.2.1: Secure transmission | • §164.312(e)(2): Transmission security | • **Art. 12** (Backup/recovery) |
| **API Security** | • **6.2.4** (protect against software attacks including APIs),<br>• **8.6.1** (system account management — acceptable),<br>• **11.6.1** (detect unauthorised changes to security-impacting content) | • §164.312(e)(1): Transmission integrity<br>• §164.312(a)(1): Access governance | • **Art. 9(4)** (network security, encryption in transit),<br>• **Art. 28** (third-party risk key principles) |
| **Continuous Monitoring** | • **10.4.1.1** (automated audit log review),<br>• **10.2.1** (audit log capture)<br>• **12.10.1**: Incident response<br>• **11.5.1**: File integrity monitoring | • §164.308(a)(1)(ii)(D): Activity reviews<br>• §164.312(b): Audit controls | • **Art. 10** (Detection), • **Art. 17** (Incident management),<br>• **Art. 19** (Reporting) |

**Table A2: Maturity Assessment Matrix with Regulatory Alignment**

| Pillar | Phase 1: Immediate Threat Neutralization (PCI DSS) | Phase 2: Core Trust Validation (HIPAA) | Phase 3: Advanced Safeguards (DORA) |
|---|---|---|---|
| **Runtime Protection** | PCI 6.2.3/6.2.4: Secure development and code review | — | DORA Art. 9: Dynamic RASP |
| **Device Trust** | PCI 8.4.2: MFA for CDE access | — | DORA Art. 9: Protection and prevention |
| **Identity Assurance** | — | HIPAA §164.312(d): Biometric authentication | DORA Art. 9: Authentication controls |
| **Data Protection** | PCI 3.5.1: Static data encryption | HIPAA §164.312(e)(2): Encrypted transmissions | DORA Art. 9(4)(b): Encryption measures |
| **Continuous Monitoring** | — | HIPAA §164.308(a)(6): Security incident procedures. | DORA Art. 10: Detection |

**Strategic Implications**

1. **Unified Compliance:** The framework consolidates overlapping controls (e.g., encryption under PCI DSS 3.5.1, HIPAA §164.312(a)(2)(iv), and DORA Art. 9(4)(b)).
2. **Risk-Based Sequencing**:
    - Phase 1 prioritizes PCI DSS requirements for payment apps.
    - Phase 2 addresses HIPAA's focus on data integrity and authentication.
    - Phase 3 aligns with DORA's advanced resilience mandates.
3. **Auditability**: Each pillar provides documented evidence for:
    - PCI DSS ROC (Report on Compliance)
    - HIPAA Security Risk Assessments
    - DORA ICT Risk Management Framework submissions.

The extended mapping to PCI DSS 4.0, HIPAA, and DORA demonstrates how the Zero Trust Mobile framework transcends sectoral boundaries. By embedding runtime-centric controls into regulatory workflows, organisations can simultaneously mitigate mobile threats and achieve compliance efficiency. Future collaborations should prioritize cross-standard control harmonization to reduce implementation overhead.